\documentclass[10pt,oneside,a4paper,twocolumn,notitlepage,fleqn]{article}
\usepackage{graphicx,epsfig}
\usepackage{float,url}
\usepackage{amsmath}
\usepackage{bbm}
\usepackage{amsfonts}
\usepackage{amssymb}

\newcommand{\figuremacroW}[4]{
	\begin{figure}[htbp]
		\centering
		\includegraphics[width=#4\textwidth]{#1}
		\caption[#2]{\textbf{#2} - #3}
		\label{#1}
	\end{figure}
}

\newtheorem{theorem}{Theorem}[section]

\makeatletter
\renewcommand\section{\@startsection{section}{2}{0cm}{-1.5ex plus -.1ex minus
 -.2ex}{1.5ex}{\centering\normalsize\sc}}
\renewcommand\subsection{\@startsection{subsection}{2}{0cm}{-1.5ex plus -.1ex minus
 -.2ex}{1.5ex}{\normalsize\it}}

\makeatother

\begin{document}
\twocolumn[
\begin{@twocolumnfalse}

\raggedleft{{\normalsize{\textmd{ISSC 2011, Trinity College Dublin, June 23--24}}}}
\rule{\textwidth}{0.3mm}
\vspace{5mm}
\begin{center}
\begin{huge}
\textmd{A Variational Bayes Approach to Decoding in a Phase-Uncertain Digital Receiver}
\end{huge}
\protect\\\vspace{5mm} 
\large\textbf{Arijit Das$^{\rm \dag}$ and Anthony Quinn$^{\rm *}$}

\vspace{5mm}
\textit{Department of Electronic and Electrical Engineering\\
Trinity College Dublin}
\protect\\\vspace{1.5mm}
%
E-mail: $^{\rm \dag}$\texttt{dasa@tcd.ie} \protect\hspace{20mm} $^{\rm *}$\texttt{aquinn@tcd.ie}

\vspace{10mm}  
\rule{\textwidth}{0.3mm} 
\vspace{3mm}
\begin{minipage}{0.9\textwidth}
\vspace{2mm}
\small{\hspace{5mm}\textit{Abstract} --- \textbf{This paper presents a Bayesian approach to symbol and phase inference in a phase-unsynchronized digital receiver.  It primarily extends \cite{Quinn2011} to the multi-symbol case,  using the variational Bayes (VB)  approximation to deal with the  combinatorial complexity of the phase inference in this case. The work provides a fully Bayesian extension of the EM-based framework underlying current turbo-synchronization methods, since it induces a von Mises prior on the time-invariant phase parmeter. As a result, we  achieve tractable iterative algorithms with improved robustness  in  low SNR regimes, compared to the current EM-based approaches. As a corollary to our analysis we also discover the importance of prior regularization in elegantly tackling the significant problem of phase ambiguity. 
}}

\vspace{1mm} \small{\hspace{5mm}}
\vspace{1mm}
\end{minipage}
\rule{\textwidth}{0.3mm}\vspace{3mm}
\end{center}
\end{@twocolumnfalse}
]

\section{Introduction}
Communication Systems face uncertainties of the transmitter and receiver oscillators, the noisy dispersive nature of the channel, among others for transferring messages from the transmitter to the receiver \cite{Shannon1984}. Incorrect alignment of the transmitter and receiver oscillators shows itself as a rotation of the received data on the complex plane and renders decoding impossible. The process of aligning these oscillators is known as synchronization and is an important aspect in the process of decoding \cite{Goldsmith2005}.  

Current state of the art in dealing with this problem is a joint iterative estimation of the phase and the symbol sequence, in an algorithm known as turbo-synchronization \cite{Herzet2007}. This algorithm is based on the framework of an EM-algorithm where probability distribution on the symbol decisions are used recursively to improve the estimate of phase, and vice-versa. This use of probability distributions over the possible symbols while making a decision, defined as a "soft" decision, was the novelty in this approach \cite{Herzet2007}. The algorithm could also be defined under other frameworks, such as one based on the sum-product algorithms \cite{Herzet2010}, \cite{Jordan1999}.

The EM-algorithm however, considers certainty-equivalence measures on the parameter values which lead to the loss information about their statistical uncertainty. This work aims to generalize this framework into a fully Bayesian treatment, which is flexible enough to "learn" the varied channel conditions along with the uncertainties associated with the modelled parameters. This was however partially achieved in \cite{Nissila2009}, wherein the authors used a uniform prior over the phase. Good Bayesian learning models take into consideration all the statistical information, the data has to offer and therefore advice better decisions under adverse conditions. 

This paper presents the advantage of using a generalization of the uniform distribution on the unit circle, the Von-Mises distribution in resolving phase ambiguity. Literature consists of either using a huge number of pilot symbols or modified likelihood estimators \cite{Herzet2010} which are generally inefficient with respect to power and time, to deal with the problem. In the course of our analysis, we find a natural elegant solution to the problem, which does not add any computational burden to the synchronization routine and exalts the importance of prior regularization in Bayesian Analysis. 

As constellation alphabets belong to finite fields, the receiver data models are inevitably mixture distributions which are not in the exponential family. Finite dimensional sufficient statistics \cite{Hipp1974} and conjugate prior distributions \cite{Diaconis1979} are a unique characteristic to the exponential family. They are important in devising recursive learning algorithms. Unfortunately as this is not the case, it leads to models with substantially expensive inference procedures. A deterministic approximation method, known as Variational Bayesian (VB) Method \cite{Smidl2006a}, is applied in this paper to deal with the problem. 

The rest of the paper is structured as follows. Section II defines the system model encompassing unknown phase and symbol sequence. Section III describes the single symbol transmission inference. Section IV describes the combinatorial complexity, the Variational Bayes Method and how it is applied to the Multi-Symbol transmission inference. In section V, numerical results for the performance of the described receiver and its comparisons with the current state of the art are provided. Section VI discusses the results and its implications. Section VII presents the concluding remarks and future course of actions.
\section{System Model}
Structured redundancy is usually added in the transmitted data \cite{Goldsmith2005}, in a process known as coding, to protect from the noisy nature AWGN channel. Equal partitions of the binary message stream formed from uniformly sampled and quantized analog signals, are bijectively mapped to a complex symbol $a = [a_1,... a_{m}, ... a_M]'$ from a constellation alphabet $\mathcal{A}$ of size M. These symbols are then modulated onto a carrier signal which can be sent through a physical channel. On the receiver side, the reverse operation called demodulation attempts to decode or identify the transmitted symbol sequence, which are now corrupted by noise \cite{Herzet2007}.

We observe the data vector $x_{i} \in \mathbb{C}^{n}$ during every symbol period with $i \in {1,... K}$ with $\textbf{x}_{i} = [x_{1}', x_2', ...x_i']'$ and assume the prior probabilities corresponding to the symbol vector $a$ to be $\alpha = [\alpha_{1}, ... \alpha_{M}]'$. $\phi$ represents the phase difference of the received signal to the transmitted, with noise $r$ in an AWGN channel \cite{Quinn2011}. Ease of inferential analysis dictates the augmentation of the model using a latent variable. Let it be $l_{i} = [l_{1,i}, ... l_{M,i}]' \in \left\{ \epsilon_{M}(1), ... \epsilon_{M}(M) \right\}$, where $\epsilon_{M}(m) = [\delta(m-1),... \delta(m-M)]'$ and $\delta$ is the Kronecker Delta. Hence $l_{i}$ is a vector pointer to the component active at the $i^{th}$ symbol period. The data generation model for such a symbol transfer in a single symbol period can be defined as:
\begin{align}
	f(x_{i} | \phi, r, \alpha) &= \sum_{l_{i}}{f(l_{i}|\alpha)f(x_{i} | a, \phi, r, l_{i})} \\
	f(x_{i} | \phi, r, l_{i}) &\equiv \mathcal{CN}_{x} \left(l_{i}'age^{j \phi}, r \right)\\
	f(l_{i}|\alpha) & \equiv \mathcal{M}u_{l_{i}}\left( 1, \alpha \right) = l_{i}' \alpha 
\end{align}
where $\mathcal{CN}$: Complex Normal and $\mathcal{M}u$: Multinomial distributions, $\phi \in [-\pi,\pi]$, $r \in \left(0, \infty \right)$, $g_{i} = [s_{1}e^{j \omega},..., s_{n}e^{j \omega n}]'$ with $s_i \in \left\{1,2 ... n \right\}$ as the pulse sequence and $\omega$ the carrier digital frequency which are assumed to be known. In this analysis we assume known SNR, based on which we can calculate the noise variance, $r$.
\subsection{Prior Distribution to Phase: Von-Mises Distribution}   
As we will see later, the VB approximation on the joint distributions lead to marginal distributions of the phase from the exponential family, hence the use of conjugate priors remains important. Von-Mises distribution \cite{Shmaliy2005}, \cite{Khatri1977} is a useful distribution over the unit circle defined from $[-\pi,\pi]$ on the real line, which serves as a conjugate prior to the phase for our conditional data model. It has one complex parameter, which serves as both location and concentration parameter for the distribution and has been used in varied applications on directional data.
\begin{align}
	f(\phi|\kappa_{0}) \equiv \mathcal{M}(\kappa_{0}) = \frac{1}{2 \pi I_{0}(|\kappa_{0}|)} e^{\left( \Re\left\{\kappa_{0}e^{-j \phi}\right\}\right)} 
\end{align}
where $\kappa_{0} \in \mathbb{C}$. In the signal processing literature this distribution is also known as the Tikhonov distribution \cite{Shmaliy2005} and has been previously used to model the phase error from the estimated phase using a phase-locked loop (PLL). In this work Von-Mises/Tikhonov distribution is considered to be the prior distribution of the phase in a Bayesian framework which was explicitly introduced in \cite{Quinn2011}. As the Von-Mises/Tikhonov distribution is confined to the unit circle, means and variances outside $[-\pi, \pi]$ are undefined. Hence circular moments are used which are defined as:
\begin{align}
	& \mathbb{E}[\phi] = \angle \kappa_{0} \\
	& Var_{cir}[\phi] = 1 - \mathbb{E}[\cos{(\phi - \mathbb{E}[\phi])}] \nonumber \\
	& \ \ \ \ \ \ \ \ \ \ \ \ \ \ = 1 - \frac{I_{1}(|\kappa_{0}|)}{I_{0}(|\kappa_{0}|)}
\end{align} 
Limiting behaviour of this distribution under various conditions of its hyperparameter are as follows
\begin{align}
	\lim_{|\kappa_{0}| \to \infty} f(\phi | \kappa_{0}) &= \mathcal{N}_{\phi}\left(\angle{\kappa_{0}},\frac{1}{|\kappa_{0}|}\right) \\
	\lim_{|\kappa_{0}| \to 0} f(\phi | \kappa_{0}) &= \mathcal{U}_{\phi}([-\pi,\pi])
\end{align}
where $\mathcal{N}$ is the normal distribution and $\mathcal{U}$ is the uniform distribution.
\section{Single Symbol Transmission}
Now we present the inference procedure for the single symbol transmission, where exact inference was shown to be possible in \cite{Quinn2011}. This analysis forms the basis for our generalization into the multi-symbol cases, wherein the Variational Bayes Method is applied. The data generation model in this case is taken to be
\begin{align}
	f(x ,l,\phi | g, \kappa_0, \alpha, \beta_{0}, a) = f(x | a, \phi, r, l)f(l|\alpha)f(\phi | \kappa_0)  
\end{align}
where
\begin{align}
	&f(\phi | \kappa_0) \equiv \mathcal{M}(\kappa_0)\\
	&f(l|\alpha)  \equiv \mathcal{M}u_{l_{t}}\left( 1, \alpha \right) =  l'\alpha\\
	&f(\textbf{x} | a, \phi, r, l) \equiv \mathcal{CN}_{\textbf{x}} \left(l'age^{j \phi}, r \right)
\end{align}
\figuremacroW{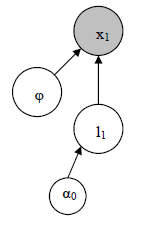}{Graphical Model}{Single-symbol model in the case of uncertain phase}{0.15}
The marginal posterior distributions for phase and the symbol can be easily calculated in this case and are given in \cite{Quinn2011} to be
\begin{align}
f(\phi | \textbf{x}) & =  \sum_{l} p_{l} \mathcal{M}(\kappa_{l}) \\
f(l | \textbf{x}) & =  \mathcal{M}u_{l}\left(1, p_{l}\right) 
\end{align}
where 
\begin{align}
\kappa_{l} & = \kappa_0 + \frac{2}{r}(l'ag)^H\textbf{x} \\
& = \kappa_0 + \frac{2}{r} (l'a)^{H} \sum_{k=1}^{n} s_{k}x_{k}e^{-j \omega k} \\
p_{l} & \propto (l'\alpha) e^{\left(-\frac{1}{r}(l'ag)^H(l'ag)\right)} I_{0}(|\kappa_{l}|)
\end{align}
The mean value of $\phi$ for each component is given by $\angle \kappa_{l}$. Also $\kappa_{l}$ can be characterized as a Distrete Time Fourier Transform on the incoming data, which is intuitive since $\kappa_{l}$ is a sufficient statistic to the phase $\phi$. Here however we get more information than from a simple DTFT, namely about the confidence over the expected phase. 

Another important observation about $\kappa_{l}$ can be seen with the following algebraic expansion:
\begin{align}
	|\kappa_{l}| & = [\kappa_{l}\overline{\kappa_{l}}]^{1/2} \\
	& = [\frac{4}{r^{2}}(l'ag)^H\textbf{x}\overline{(l'ag)^H\textbf{x}} \nonumber \\ 
	& + \kappa_0\overline{\kappa_0} +  \frac{2}{r}\kappa_0\overline{(l'ag)^H\textbf{x}} + \frac{2}{r}\overline{\kappa_0}(l'ag)^H\textbf{x}]^{1/2}
\end{align}
%
From the above equation we can see that when $\kappa_{0} = 0$ (Uniform Prior), all except the first term vanishes. That implies then that when the prior is a uniform distribution, the decoding distribution is invariant to the symbols on the same radius in a QAM constellation, which is the origin of phase ambiguity. However a non-zero $\kappa_0$ makes $\kappa_{l}$ depend on the value of the symbols and hence results in a unique MAP from the decoding distribution.    
\section{Multi-Symbol Transmission}
Real world communication systems involve multiple symbol periods, which are observed sequentially and a decoding distribution for the symbols is the desired output. The joint data model under the independent symbol sequence assumption (reasonable in turbo processing where pseudorandom inter-leavers are typically employed \cite{Nissila2009}) is then 
\begin{align}
	f(\textbf{x}_{K} |\phi, r, \alpha) = \prod_{i=1}^{K}{ \sum_{l_{i}}{f(l_{i}|\alpha)f(x_{i} | a, \phi, r, l_{i})}} 
\end{align}
As a consequence of this product over a summation, $M^K$ term mixture model appears after $K$ time periods, a combinatorial explosion which leads to intractable exact inference.
\subsection{Variational Bayes Method}
A deterministic approximation, known as the Variational Bayes Method is employed to deal with this computational intractability. The Method aims at minimizing the Kullback-Leibler Divergence between the candidate approximation and the original joint distribution. This divergence can be interpreted as an information difference \cite{Wainwright2008} between the two distributions. The intuition then is to get a functional form of the posterior marginals, so as to minimize the information lost due to the forced independence \cite{Smidl2006}. These methods have found many applications in machine learning, artificial intelligence and more recently in signal processing \cite{Nissila2009}. The essential result which forms the basis of this method is as follows:
\begin{theorem}
Variational Bayes Method \cite{Smidl2006a}: Let $f(\theta|\textbf{x})$ be the posterior distribution of multivariate parameter, $\theta$. The latter is partitioned into q sub-vectors of parameters or q-nodes:
\begin{align}
	\theta = [\theta_{1}',\theta_{2}',...\theta_{q}']'
\end{align}
Let $\breve{f}(\theta|\textbf{x})$ be an approximate distribution restricted to the set of conditionally independent distributions for $\theta_{1}$,$\theta_{2}$,...$\theta_{q}$:
\begin{align}
	\breve{f}(\theta|\textbf{x})= \prod_{i=1}^{q} \breve{f}(\theta_{i}|\textbf{x})
\end{align}
Then, the minimum of Kullback-Leibler divergence $\mathbf{KL}\left( \breve{f}(\theta|\textbf{x}) || f(\theta|\textbf{x}) \right)$ is reached for
\begin{align}
	\tilde{f}(\theta_{i}|\textbf{x}) & \propto e^{ \left( \mathbb{E}_{\tilde{f}(\theta_{/i}|\textbf{x})}\left[ \ln \left( f(\theta,\textbf{x}) \right)\right] \right)}
\end{align}
where $i = 1,...,q$ and $\theta_{/i}$ in $\theta$, and $\tilde{f}(\theta_{/i}|\textbf{x}) = \prod_{j=1,j \neq i}^{q} \tilde{f}(\theta_{j}|\textbf{x})$. Now $\tilde{f}(\theta|\textbf{x})$ is the VB-aproximation, and $\tilde{f}(\theta_{i}|\textbf{x})$ are the VB-marginals.
\end{theorem}
\subsection{Offline VB Approximation}
Consider the situation, where we observe a batch of symbol transfer periods together and are required to infer the marginal distributions for all the transferred symbols and the phase parameter. Prior independence of the symbols is assumed. However it should be noted, that aposteriori the symbols are independent conditional on $\phi$. Expanding the joint probability, over all unknown parameters, we have
\begin{align}
f(\textbf{x}_K , & \textbf{l}_{K},\phi| \kappa_0, \alpha, r) \nonumber \\
& = f(\phi | \kappa_0) \prod_{i=1}^{K} f(x_{i} | a, \phi, r, l_{i})f(l_{i}|\alpha) 
\end{align}
where
\begin{align}
	&f(\phi | \kappa_0) \equiv \mathcal{M}(\kappa_0)\\
	&f(l_{i}|\alpha)  \equiv \mathcal{M}u_{l_{i}}\left(1, \alpha\right) = l_{i}' \alpha \\
	&f(x_{i} | a, \phi, r, l_{i}) \equiv \mathcal{CN}_{x_{i}} \left(l_{i}'ag_{i}e^{j \phi}, r \right)
\end{align}
\figuremacroW{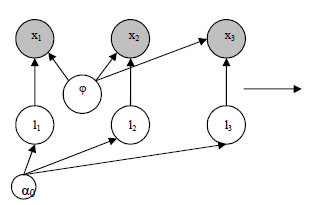}{Graphical Model}{Multi-symbol model in the case of uncertain phase}{0.40}
From the above joint distribution, we find the posterior distributions for the label field and the phase. Integrating out the label field entails a $M^K$ summation, making it intractable \cite{Smidl2006}. We therefore resort to using Theorem II.1 and derive the approximate posterior distributions of the parameters as
\begin{align}
	\tilde{f}(\phi|\textbf{x}_{K}) & \propto e^{ \left( \mathbb{E}_{\tilde{f}(\textbf{l}_{K}|\textbf{x}_{K})} [\ln{f(\textbf{x}_K , \textbf{l}_{K},\phi| \kappa_0, \alpha, r)}]\right)} \nonumber \\
	& = \mathcal{M}_{\phi}(\kappa_{\textbf{l}_{K}}) 
\end{align}
\begin{align}	
	\tilde{f}(l_{i}|\textbf{x}_{K}) & \propto e^{ \left( \mathbb{E}_{\tilde{f}(\phi|\textbf{x}_{K})\tilde{f}(\textbf{l}_{K \backslash i}|\textbf{x}_{K})} [\ln{f(\textbf{x}_K , \textbf{l}_{K},\phi| \kappa_0, \alpha, r)}]\right)} \nonumber \\
	& = \mathcal{M}u_{l_{i}}(1, p_{l_{i}}) 
\end{align}
with the shaping parameters as
\begin{align}
	& \kappa_{\textbf{l}_{K}} = \kappa_0 +  \frac{2}{r} \sum_{i=1}^{K} (\hat{l}_{i}'ag_{i})^{H}x_{i} \\
	& p_{l_{i}} \propto (l_{i}'\alpha) e^{\left(- \frac{1}{r} (l_{i}'ag_{i})^{H}(l_{i}'ag_{i}) + \frac{2}{r}\delta_{i} \right)}
\end{align}
and the associated VB moment as
\begin{align}
	\delta_{i} & = \mathbb{E}_{\tilde{f}(\phi|\textbf{x}_{K})}\left[ \Re\left\{(l_{i}'ag_{i})^Hx_{i} e^{-j \phi}\right\} \right] \nonumber \\ 
	& = \Re \left\{(l_{i}'ag_{i})^H x_{i} e^{-j \phi_{\kappa_{\textbf{l}_{K}}}} \right\}\frac{I_{1}(|\kappa_{\textbf{l}_{K}}|)}{I_{0}(|\kappa_{\textbf{l}_{K}}|)}
\end{align}
$\kappa_{\textbf{l}_{K}}$ again can be characterized as a DTFT. Important subsets of this method discussed in the literature include \cite{Nissila2009} where the authors have assumed an uniform prior to the phase which implies $\kappa_{0}=0$. Further if we make the ratio $$\frac{I_{1}(|\kappa_{\textbf{l}_{K}}|)}{I_{0}(|\kappa_{\textbf{l}_{K}}|)} = 1$$ we get the EM-algorithm based turbo-synchronization routine as discussed in \cite{Herzet2007}.
\subsection{Online VB Approximation}
In this case we observe each symbol transfer period sequentially and are required to infer the marginal distributions for each transferred symbol and phase parameter, after each symbol period. Again prior independence of the symbols is assumed. To deal with the increasing combinatorial complexity after each time step we conflate the posterior distribution of $\phi$ which is a mixture of $M$ Von-Mises components using Variational Bayes Method in to a single Von-Mises posterior. The associated sufficient statistic can then be used as the prior for the next symbol period. Expanding the joint probability, over unknown parameters after each time step, we have
\begin{align}
	f(x_{t},l_{t},\phi| \textbf{x}_{t-1}) = f(x_{t} | a, \phi, r, l_{t})f(l_{t}|\alpha) \tilde{f}(\phi|\textbf{x}_{t-1})
\end{align}
where
\begin{align}
	&f(\phi | \textbf{x}_0) = f(\phi | \kappa_0) \equiv \mathcal{M}(\kappa_0)\\
	&f(l_{t}|\alpha)  \equiv \mathcal{M}u_{l_{t}}\left(1, \alpha\right) = l_{t}' \alpha \\
	&f(x_{t} | a, \phi, r, l_{t}) \equiv \mathcal{CN}_{x_{t}} \left(l_{t}'ag_{t}e^{j \phi}, r \right)
\end{align}
On equation (13) from Section III of the posterior distribution of the phase, we apply the VB Method and derive the approximate conflated posterior distribution as 
\begin{align}
	\tilde{f}(\phi |\textbf{x}_{t-1}) & \propto e^{\left( \mathbb{E}_{\tilde{f}(l_{t-1}|x_{t-1})} [\ln{f(x_{t-1},l_{t-1},\phi| \textbf{x}_{t-2}, \kappa_0, \alpha, r)}]\right)} \nonumber \\
	& = \mathcal{M}_{\phi}(\kappa_{\textbf{l}_{t-1}}) \\
\end{align}
with the shaping parameter as
\begin{align}
	\hat{\kappa}_{\textbf{l}_{t-1}} &= \hat{\kappa}_{\textbf{l}_{t-2}} +  \frac{2}{r}(\hat{l}_{t-1}'ag_{t-1})^{H}x_{t-1} \\
	 p_{l_{t-1}} &\propto (l_{t-1}'\alpha) e^{\left(-\frac{1}{r}(l_{t-1}'ag)^H(l_{t-1}'ag)\right)} I_{0}(|\kappa_{\textbf{l}_{t-1}}|) \\
	 \hat{l}_{t-1} &=  \mathbb{E}[l_{t-1}]
\end{align}
Using this shaping parameter as initialization for the following symbol period we get from the single symbol transmission case:
\begin{align}
f(\phi | \textbf{x}_{t}) & =  \sum_{l} p_{\textbf{l}_{t}} \mathcal{M}(\kappa_{\textbf{l}_{t}}) \\
f(l_{t} | \textbf{x}_{t}) & =  \mathcal{M}u_{l_{t}}(1, p_{l_{t}}) 
\end{align}
where 
\begin{align}
& \kappa_{\textbf{l}_{t}} = \hat{\kappa}_{\textbf{l}_{t-1}} + \frac{2}{r}(l_{t}'ag_{t})^{H}x_{t} \\
& p_{l_{i}} \propto (l_{i}'\alpha) e^{\left(-\frac{1}{r}(l_{i}'ag)^H(l_{i}'ag)\right)} I_{0}(|\kappa_{\textbf{l}_{t}}|)
\end{align}
$\kappa_{\textbf{l}_{t}}$ again can be characterized as a DTFT as earlier.
\section{Simulation}

\figuremacroW{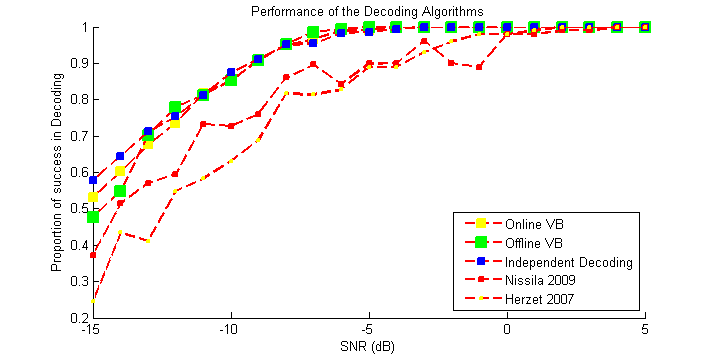}{}{Performance of the VB-based decoding algorithms in the multi-symbol case}{0.52}

The above figure describes the proportion of successful identifications of the symbol transferred for various values of SNR, using the methods developed in this paper. In each batch 20 symbols are transmitted through an AWGN channel. In the low-SNR regime, from -15dB to 5dB we see that the Bayesian algorithms are significantly better. Independent decoding algorithm uses the exact decoding algorithm over each subsequent symbol period ignoring the previous periods. This method turns out to be the fastest and most accurate for this simulation. The Online VB and the Offline VB approximations provide almost as good results, though they would be more robust in a more general setting. We can also clearly see the performance of \cite{Nissila2009} and \cite{Herzet2007} is significantly lower with respect to the decoding success. It should be noted, to deal with the problem of phase ambiguity for these two methods, 5 pilot symbols are used in front of the original 20.

\section{Discussion}

We get two main results from the analysis performed in this paper. Primarily we are able to extend the EM-Algorithm based synchronization into a fully Bayesian synchronization procedure. This can form the basis for future generalizations and analysis of even more complicated scenarios. Secondly, as a corollary to our analysis we are able to resolve the problem of phase ambiguity.

As shown in the  Section~III, setting $\kappa_{0}=0$ renders $\kappa_{l}$ independent of the expected phase $\mathbb{E}[\phi]$, which thereby results in the apparent ambiguity in the decoding distribution. The notion of non-zero $\kappa_{0}$ seems quite inconsequential, but a concentrated prior however informs the presence of a unimodal phase distribution and regularizes the observation model which is intrinsically rotationally invariant for QAM. It must be noted however, that as the number of incoming observations increases the sampling distribution of the MAP symbol becomes relatively invariant to the value of $\kappa_0$, as long as $\kappa_0>0$.

Computationally it presents no overhead, as it involves only the addition of a constant. In contrast past approaches like Differential Coding \cite{Goldsmith2005}, angle differential QAM scheme \cite{Hwang2008} and the use of pilot symbols \cite{Herzet2007} \cite{Herzet2010}, among others which have been proposed to deal with this rotational invariance are visibly more complex and lack the simplicity of this observation.  

In \cite{Herzet2007} the authors use the EM-Algorithm to deal with phase synchronization. This essentially implies using the sifting property of the Kronecker Delta function and thus using the expectations of the parameters in the associated functions rather than the expectations of the functions of the parameters itself. Significantly lower accuracy of the decoding distribution in the low-SNR regime is observed which converges in performance with increasing SNR.

\section{Conclusions}
In this paper we have been able to extend the EM framework for tackling the phase-synchronization problem, to a fully Bayesian VB based algorithm wherein as a virtue of its modelling ability the problem of phase ambiguity does not arise. This makes the concept of using "soft" information truly complete as no statistical information is lost at any point in the iterations. Further the basic framework developed motivates further relaxations into more general channels, such as those with multi-path fading, like Rician Fading \cite{Goldsmith2005}. In future work we would also be relaxing the condition on known noise parameter and deal with it in a Bayesian framework.

\section{Acknowledgments}
This research was supported by Science Foundation Ireland, grant 08/RFP/MTH1710
\begin{small}


%

\end{small}


\begin{thebibliography}{10}

\bibitem{Diaconis1979}
{\sc Persi Diaconis and Donald Ylvisaker}.
\newblock {\bf Conjugate Priors for Exponential Families}.
\newblock {\em The Annals of Statistics}, {\bf 7}(2):pp. 269--281, 1979.

\bibitem{Goldsmith2005}
{\sc Andrea Goldsmith}.
\newblock {\em Wireless Communications}.
\newblock Cambridge University Press, New York, NY, USA, 2005.

\bibitem{Herzet2007}
{\sc C.~Herzet, N.~Noels, V.~Lottici, H.~Wymeersch, M.~Luise, M.~Moeneclaey,
  and L.~Vandendorpe}.
\newblock {\bf Code-Aided Turbo Synchronization}.
\newblock {\em Proceedings of the IEEE}, {\bf 95}(6):1255 --1271, june 2007.

\bibitem{Herzet2010}
{\sc C.~Herzet, K.~Woradit, H.~Wymeersch, and L.~Vandendorpe}.
\newblock {\bf Code-Aided Maximum-Likelihood Ambiguity Resolution Through
  Free-Energy Minimization}.
\newblock {\em Signal Processing, IEEE Transactions on}, {\bf 58}(12):6238
  --6250, Dec 2010.

\bibitem{Hipp1974}
{\sc Christian Hipp}.
\newblock {\bf Sufficient Statistics and Exponential Families}.
\newblock {\em The Annals of Statistics}, {\bf 2}(6):pp. 1283--1292, 1974.

\bibitem{Hwang2008}
{\sc Jeng-Kuang Hwang, Yu-Lun Chiu, and Chia-Shu Liao}.
\newblock {\bf Angle differential-QAM scheme for resolving phase ambiguity in
  continuous transmission system}.
\newblock {\em Int. J. Commun. Syst.}, {\bf 21}:631--641, June 2008.

\bibitem{Jordan1999}
{\sc Michael~I. Jordan}.
\newblock {\bf An introduction to variational methods for graphical models}.
\newblock In {\em Machine Learning}, pages 183--233. MIT Press, 1999.

\bibitem{Khatri1977}
{\sc C.~G. Khatri and K.~V. Mardia}.
\newblock {\bf The Von Mises-Fisher Matrix Distribution in Orientation
  Statistics}.
\newblock {\em Journal of the Royal Statistical Society. Series B
  (Methodological)}, {\bf 39}(1):pp. 95--106, 1977.

\bibitem{Nissila2009}
{\sc Mauri Nissila and Subbarayan Pasupathy}.
\newblock {\bf Adaptive iterative detectors for phase-uncertain channels via
  variational bounding}.
\newblock {\em Trans. Comm.}, {\bf 57}:716--725, March 2009.

\bibitem{Quinn2011}
{\sc Anthony Quinn, Jean-Pierre Barbot, and Pascal Larzabal}.
\newblock {\bf The Bayesian Inference of Phase}.
\newblock In {\em Proceedings of IEEE International Conference on Acoustics,
  Speech and Signal Processing, 2011}, ICASSP'11, 2011.

\bibitem{Shannon1984}
{\sc C.E. Shannon}.
\newblock {\bf Communication in the presence of noise}.
\newblock {\em Proceedings of the IEEE}, {\bf 72}(9):1192 -- 1201, sept 1984.

\bibitem{Shmaliy2005}
{\sc Yuriy~S. Shmaliy}.
\newblock {\bf Von Mises/Tikhonov-based distributions for systems with
  differential phase measurement}.
\newblock {\em Signal Processing}, {\bf 85}(4):693 -- 703, 2005.

\bibitem{Smidl2006}
{\sc V.~Smidl and A.~Quinn}.
\newblock {\bf The Variational Bayes Approximation In Bayesian Filtering}.
\newblock In {\em IEEE International Conference on Acoustics, Speech and Signal
  Processing, 2006}, ~{\bf 3}, page III, May 2006.

\bibitem{Smidl2006a}
{\sc Vaclav Smidl and Anthony Quinn}.
\newblock {\em The Variational Bayes Method in Signal Processing}.
\newblock Springer, Dublin, Ireland, 2006.

\bibitem{Wainwright2008}
{\sc Martin~J. Wainwright and Michael~I. Jordan}.
\newblock {\bf Graphical Models, Exponential Families, and Variational
  Inference}.
\newblock {\em Foundations and Trends® in Machine Learning}, {\bf 1}:1--305,
  2008.

\end{thebibliography}
\end{document}